\documentclass[ preprint,  amsmath,amssymb,  aip, ]{revtex4-1}
\usepackage{graphicx}
\usepackage{color}
\usepackage{dcolumn}
\usepackage{bm}
\usepackage{amsmath}
\usepackage{amsfonts}
\usepackage{amssymb}%

\providecommand{\U}[1]{\protect\rule{.1in}{.1in}}
\begin{document}

\title{Manuscript Title
\\with Forced Linebreak}%

\begin{abstract}
We investigated the effect of an electric field on the interface magnetic anisotropy of a thin MgO/Fe/MgO layer using density functional theory. The
perpendicular magnetic anisotropy energy (MAE) increases not only under electron depletion but also under some electron accumulation
conditions, showing a strong correlation with the number of electrons on the interface Fe atom. The reverse variation in the MAE under the electric
field is ascribed to novel features on the charged interface, such as electron leakage. We discuss the origin of the variation in terms of the
electronic structures.
\end{abstract}

\title{
Possible origin of nonlinear magnetic anisotropy variation
in electric field effect in a double interface system
}

\author{Daiki Yoshikawa} \email{yoshikawa@cphys.s.kanazawa-u.ac.jp}%
\affiliation{
Graduate School of Natural Science and Technology, Kanazawa
University, Kanazawa 920-1192, Japan
}

\author{Masao Obata}
\affiliation{
Graduate School of Natural Science and Technology, Kanazawa
University, Kanazawa 920-1192, Japan
}

\author{Yusaku Taguchi}
\affiliation{
Graduate School of Natural Science and Technology, Kanazawa
University, Kanazawa 920-1192, Japan
}

\author{Shinya Haraguchi}
\affiliation{
Graduate School of Natural Science and Technology, Kanazawa
University, Kanazawa 920-1192, Japan
}
\author{Tatsuki Oda}
\affiliation{
Graduate School of Natural Science and Technology, Kanazawa
University, Kanazawa 920-1192, Japan
}

\affiliation{
Institute of Science and Engineering, Kanazawa University, Kanazawa
920-1192, Japan}

% \author{Daiki Yoshikawa$^1$, Masao Obata$^1$, Yusaku Taguchi$^1$, 
% Shinya Haraguchi$^1$, and Tatsuki Oda$^{1,2}$}
% \affiliation{
% Graduate School of Natural Science and Technology, Kanazawa
% University, Kanazawa 920-1192, Japan
% Institute of Science and Engineering, Kanazawa University, Kanazawa
% 920-1192, Japan}

% ${\rm E}$-${\rm mail: yoshikawa@cphys.s.kanazawa}$-${\rm u.ac.jp}$
% }

\maketitle

  Electric-field- (EF-) driven devices of magnetic
  materials have been a possible direction for future
  spin electronic applications for more than ten
  years\cite{Ohno2010,Ohno2000,Weisheit2007,Maruyama2009}.
  Metallic devices with an interface of insulating
  material have become a promising system. Ultralow energy
  consumption is strongly expected owing to the nonvolatile
  nature of magnetism\cite{Shiota2012}. In magnetic tunnel junctions, duplication
  of a single junction has been introduced to improve
  performance\cite{Naganuma2011}. Similarly, in a magnetic device intended to
  exploit the EF-driven change in the magnetic anisotropy,
  a proposed double interface structure has shown a large
  enhancement in the EF-induced effect on the magnetic
  anisotropy energy (MAE)\cite{Nozaki2013}. In this work, the authors
  found unusual nonlinear behavior of the EF dependence
  of the MAE. For both electron depletion and electron
  accumulation conditions at the interface of the magnetic
  metallic layer, the MAE changes to favor stability in the
  magnetic direction perpendicular to the interface plane. The
  origin of this preference in such a double interface structure
  is not yet clear.
  The establishment of nonlinear behavior in the MAE variation
  may extend the range of applications of EF-driven magnetic
  devices\cite{Suzuki2011}. A theoretical understanding of the behavior
  will accelerate development. In this work, we successfully
  explain this behavior using a realistic model with a double
  interface, while MAE variation has been investigated theoretically
  only at single interfaces\cite{Niranjan2010, Nakamura2010}.
  We used the MgO/Fe
  interface, which shows a magnetic anisotropy perpendicular
  to the interface plane. The effect of B atoms in the magnetic
  layer\cite{Nozaki2013} was neglected. This effect has been investigated in
  the literature\cite{Hotta2013, Miyajima2009}.

\begin{figure}[t]
\begin{center}
\includegraphics[width=85mm]{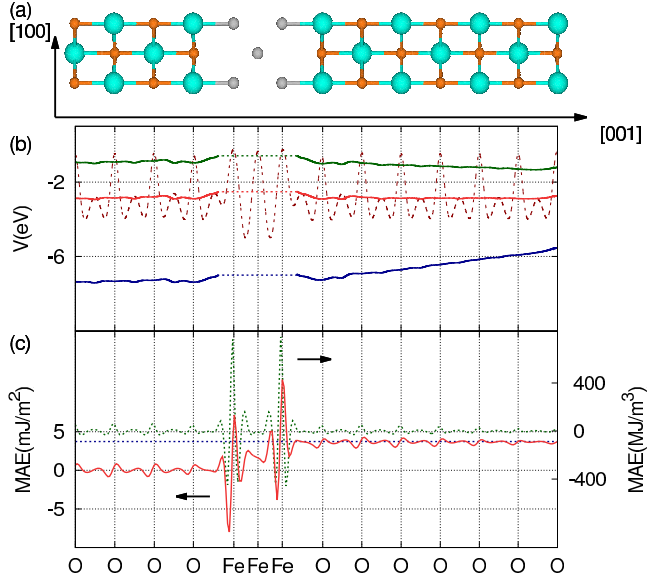}
\end{center}
 \caption{(a) Atomic configuration of slab, MgO (4 ML)/Fe (3 ML)/MgO (7 ML). (b) Averaged electrostatic potentials for electrons along the $z$-axis at $E=0$ (red thin and thick curves), $E=0.42$ V/nm (blue thick curves), and $E=-0.26$ V/nm (green thick curves). (c) MAE distribution averaged over the $xy$-plane (green dotted curve) and integrated along the $z$-axis (red thick curve). Blue dotted line specifies the MAE (3.73 mJ/m$^2$) at $E=-0.16$ V/nm.}
\label{Str_charge_map}
\end{figure}

We used a slab system, vacuum (0.79 nm)/MgO [4 atomic
     monolayers (ML)]/Fe (3 ML)/MgO (7 ML)/vacuum (0.79
     nm), in the computation [Fig. 1(a)]. The atoms are specified
     by number as Fe(1), Fe(2), ..., O(1), ..., Mg(1), ..., etc.,
     from the left-hand side of the system. The sets of left and
     right MgO layers are labeled MgO(L) and MgO(R), respectively.
     There are two MgO/Fe interfaces in the model. At
     each interface, an Fe atom was placed directly next to the O
     atom because of its stability. This system includes both edges
     of the magnetic layer, enabling us to investigate the EF effect,
     which comes from both simultaneously.
     We carried out a first-principles density functional calculation
     that uses fully relativistic ultrasoft pseudopotentials
     and a planewave basis\cite{Oda2005} by using the generalized gradient
     approximation\cite{Perdew1992}. The MAE was estimated from the total
     energy difference between the [100] ($x$-axis) and [001]
     ($z$-axis) magnetization directions, that is, ${\rm MAE}=E[100]-E[001]$
     , where [001] specifies the direction on the right-hand
     side in Fig. 1(a). We used a 32 $\times$ 32 $\times$ 1 mesh in {\bf k} point
     sampling, and the in-plane lattice constant was fixed at
     the value for the MgO layer extracted from the bulk
     ($a$ = 0.298 nm). We induced structural relaxation at zero EF
     while maintaining both the in-plane lattice constant and the
     atomic coordinates of Mg(1) and O(1) in a plane perpendicular
     to [001]. To apply the EF, we placed an effective
     screened medium (ESM)\cite{Otani2006} on the right-hand edge of the slab
     system (adjacent to the vacuum). This medium acts like an
     ideal metal, accumulating charge on its surface when the slab
     system becomes charged. Therefore, by introducing a change
     in the number of electrons (NOE) in the slab, an EF is
     imposed on the slab from the right, causing an EF to appear
     on the magnetic layer in the MgO(R) layer. Consequently,
     we could investigate the effects of the EF on the magnetic
     metallic layer sandwiched with MgO. This situation is
  thought to be similar to that in experimental work\cite{Nozaki2013}, in
  which most of the applied voltage falls across the thicker
  MgO layer corresponding to MgO(R). The external EF $E_{{\rm ext}}$
  can be estimated from the gradient of the electrostatic
  potential at the front of the ESM in the vacuum layer. The
  details of the EF application are given in a previous work\cite{Tsujikawa2009}.
  The MAE will be discussed with respect to the realistic EF $E$,
  which is obtained from $E=E_{\rm ext}/\varepsilon_{\rm r}$, where $\varepsilon_{\rm r} (=9.8)$ is the
  relative dielectric constant for MgO. EFs of 0.42, 0.21, 0.0,
  $-$0.10, $-$0.16, $-$0.21, and $-$0.26V/nm were obtained, with
  electron depletions or accumulations in the slab of $-$0.02,
  $-$0.01, 0.0, 0.005, 0.0075, 0.01, and 0.0125, respectively.
  We calculated the MAE density by analyzing the real-space
  distribution of the MAE\cite{Tsujikawa2009,Tsujikawa2011}.

    At the interfaces, the distances between the Fe and O atoms were
  estimated to be 2.22${\rm \AA}$. This value is consistent with
  experimental and theoretical values determined in these interface
  configurations\cite{Nakamura2010,Tusche2006}. The Mg atoms were
  relaxed in the direction of the Fe layer by a small displacement, such
  as 5 pm,\cite{Nakamura2010} from the O layer. This means that there
  existed an electric polarization directed in the perpendicular
  ($z$-axis) direction. The Mg displacement is a consequence of the
  electron transfer caused by orbital hybridization between Fe
  3$d(3z^2-r^2)$ and O 2$p(z)$, which is partially reduced by the
  displacement of electron densities on the Mg atom. As a result of the
  electronic structures and electric polarizations at the MgO/Fe
  interface, the electrons with energy around the Fermi level did not
  inhabit firmly stable states on Fe(1) and Fe(3). From the analyses in
  our calculation, the electron potential at the interface rose at the
  Fe and decreased at the MgO, as shown in Fig. 1(b). In this figure,
  the electrostatic potential for $E=0$ averaged within the plane normal
  to [001] is presented as a broken curve, in addition to the further
  averaged potentials (solid green, red, and blue curves) for $E=$
  $-$0.26, 0.0, and 0.42V/nm, which were obtained by averaging data in a
  period of the layer along [001]. The potential shoulder observed at
  the interface may indicate a novel property of electrons. This
  potential weakens the ability to hold electrons on the Fe; that is,
  the number of electrons on the Fe becomes sensitive to the external
  EF. In Fig. 1(c), the real-space variation in the MAE and the
  integrated distribution along the $z$-axis are presented. These curves
  help us to understand that the MgO/Fe/MgO magnetic layer/interface
  produces perpendicular magnetic anisotropy (positive
  MAE). Unfortunately, because of the smallness of the MAE variation,
  these curves could not be used to identify the EF effect. 

\begin{figure}[t]
\begin{center}
\includegraphics[width=85mm]{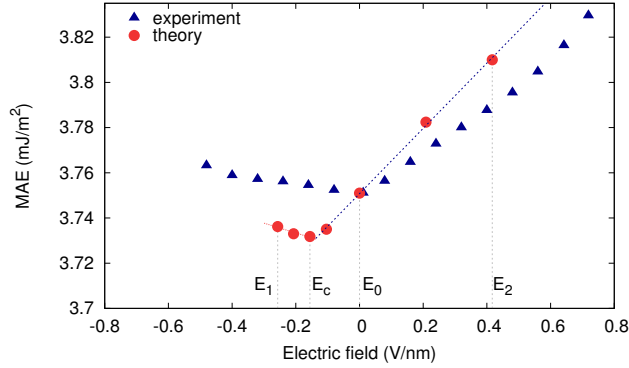}
\end{center}
\caption{
 EF variation of MAE (red circles) compared with the experimental data (blue triangles), which are adjusted to the MAE value at zero EF.
 }
\label{MAEs}
\end{figure}

  The MAEs with respect to the EF are presented and compared with experimental
  data\cite{Nozaki2013} in Fig. 2. To check an unambiguity on the number of MgO layers,
  we also estimated the MAEs in a similar system, MgO (5 ML)/Fe (3
  ML)/MgO (6 ML). The EF-induced variation in the MAE did not change
  except for small uniform increases by about 0.003 mJ/m$^2$.  The
  perpendicular magnetic anisotropies were obtained and were consistent
  with previous theoretical calculations and experimental
  observations\cite{Shimabukuro2010, Ikeda2010}. The depletion of electrons on the interface, which
  corresponds to a positive EF, increased the MAE at a rate of 144
  fJ/Vm. This rate is in good agreement with experimental data\cite{Nozaki2010,Endo2010},
  theoretical estimations\cite{Shimabukuro2010, Niranjan2010}, and, in particular, the experiment on
  the double interface (108 fJ/Vm)\cite{Nozaki2013}. The electron accumulation caused
  the MAE to decrease at the same rate as under a positive EF until EF
  $=E_{\rm c}$ and subsequently reversed the variation in the MAE. This reverse
  corresponded well to an experimental observation whose origin was not
  clarified\cite{Nozaki2013}. The variation rate at $E < E_{\rm c}$ was estimated to be $-43$
  fJ/Vm, which is comparable to the experimental value ($-24$ fJ/Vm)\cite{Nozaki2013}.

\begin{figure*}[t]
\begin{center}
\includegraphics[width=155mm]{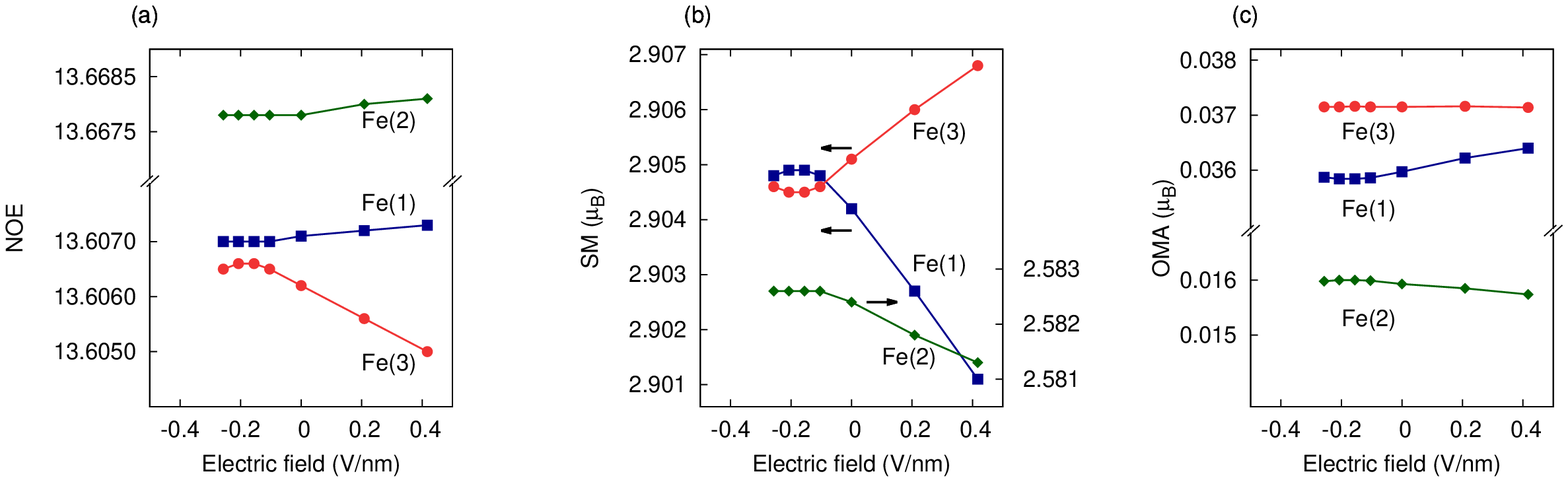}
\end{center}
\caption{
 EF variations in (a) NOE on Fe (1) (blue circles), Fe (2) (green diamonds), and Fe (3) (red squares), (b) atomic SM, and (c) OMA.}
\label{charge_spin}
\end{figure*}

  To approach the origin of the effect of the EF on the MAE variation, the
  NOE, spin magnetic moment (SM), and orbital magnetic moment anisotropy
  (OMA)\cite{Tsujikawa2012} were investigated.  The atomic quantities are presented in
  Fig. 3. The MAE has been discussed in terms of the variation in the
  NOE\cite{Nakamura2009}. As shown in Fig. 3(a), the variation in the NOE on Fe(3) was
  apparently correlated with the MAE variation shown in Fig. 2. Further,
  most of the variation in the MAE is described by a rate of $-47$ ($-40$) mJ/m$^2$
  per NOE on Fe(3) at EFs larger (smaller) than $E_{\rm c}$. There is an
  interesting feature in the variation in the NOE. When the NOE on Fe(3)
  decreases, that on both Fe(1) and Fe(2) increases, implying electron
  transfer from Fe(3) to Fe(1) and Fe(2). This contrast in the NOE
  variations enabled us to recognize the enhancement of electron
  depletion on Fe(3), highlighting the large effect of the EF on the MAE
  variation (144 fJ/Vm) at positive EFs.  Under a negative EF, the NOE
  on Fe(3) increased until $E_{\rm c}$ and then decreased. An analysis of the
  electron distribution revealed that, at $E < E_{\rm c}$, the distribution
  shifted toward the surface of MgO(R) in our model. This implies that,
  in the experimental system, electrons move into the insulating MgO(R)
  layer at negative EFs [see the negative slope of the electrostatic
  potential in Fig. 1(b)]. Note that such a density shift in our model
  never implies electrons occupying the state in the vacuum region.

  The Fe atom at the interface has a large magnetic moment ($\sim 3\ \mu_{\rm B}$), as shown
  in Fig. 3(b). This indicates a large exchange splitting ($\sim 3\ {\rm eV}$) in the
  electronic structure.  Because of this structure, a decrease (an
  increase) in the NOE usually corresponds to an increase (a decrease)
  in the SM. However, in Fig. 3(b), the EF-induced variation in the SM
  on Fe atoms may be enhanced by the spin flip in Fe. This is because
  the slope of the variation in the SM becomes steeper than that
  expected from the variation in the NOE.  According to Bruno's
  relation\cite{Bruno1989}, the OMA may reflect the behavior of the MAE. In
  Fig. 3(c), the variations imply an EF dependence of the MAE. The sum of the OMAs over Fe
  atoms (not shown) increased (decreased) under EFs in the range of $E > E_{\rm c}$ ($E < E_{\rm c}$.
  However, assuming an atomic spin-orbit coupling, the
  variation rates of the MAE were much lower than those from the total
  energy computation and experiment ($-0.2$ and $16$ fJ/Vm for $E < E_{\rm c}$ and $E > E_{\rm c}$,
  respectively).

\begin{figure}[t]
\begin{center}
\includegraphics[width=75mm]{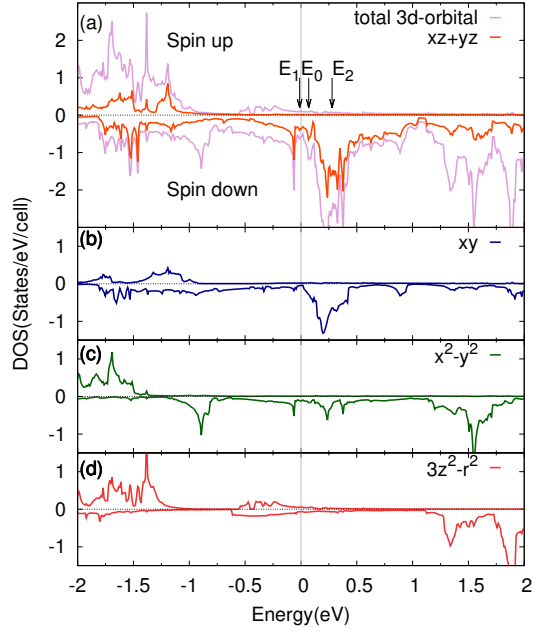}
\end{center}
\caption{
 PDOS at $E_{\rm c}$ on the 3$d$ orbitals of Fe (3). (a) Total 3$d$ orbital and $d(xz)+d(yz)$, (b) $d(xy)$, (c) $d(x^2-y^2)$, (d) $d(3z^2-r^2$). Vertical line and arrows indicate the location of the Fermi level at $E=E_{\rm c}$, \textcolor{black}{$E_{0}$($=0$),} $E_1$($=-0.26$ V/nm), and $E_2$ ($=0.42$ V/nm). Note that the EF-induced changes in the structure of the PDOS are negligible.}
\label{DOS}
\end{figure}

The projected densities of states (PDOSs) on
  Fe(3) around the Fermi energy $E_{\rm F}$ are shown for $E=E_{\rm c}$ in Fig. 4. The
  corresponding PDOSs on Fe(1) were very similar to those on Fe(3). The
  PDOSs on Fe(2) did not make any large contribution around
  $E_{\rm F}$. Interestingly, the orbitals of $d(xz)+d(yz)$ and $d(xy)$ had peaks $0.2$ eV
  above $E_{\rm F}$ at $E=E_{\rm c}$. These atomic orbitals extended in the angular
  directions along which there is no covalent bond to the
  interface. They were localized at both Fe(1) and Fe(3) and did not
  exist on Fe(2). The peaks in the PDOSs come from two-dimensional van
  Hove singularities at the flat bands (not shown) in the momentum space
  (first Brillouin zone). Such localized and dispersionless electronic
  states may play a role in capturing electrons when they are
  occupied. The location of the peaks mentioned above agrees well with
  those of the interface resonance state (IRS) observed in
  experiments\cite{Zermatten2008}. It is also interesting to see that $E_{\rm F}$ of $E=E_{\rm c}$
  existed at a dip in the total $3d$ PDOS. This dip may imply the
  appearance of a reverse point in the EF-induced MAE variation. Note
  that the $E_{\rm F}$ value determined in our calculation represents that of the
  entire system but not that of the metallic layer. Thus, when the EF
  changed toward positive (negative) values, the location of $E_{\rm F}$ shifted
  to a high (low) energy, as indicated in Fig. 4. For example, the
  location of $E_{\rm F}$ at $E=E_2$ shifted to around the large peaks as a
  consequence of the lowered electron potential [see Fig. 1(b)] at the
  magnetic layer. This did not indicate that the IRS was occupied by the
  redistributed electron. If it is possible to define an effective $E_{\rm F}$
  for the magnetic layer, it should be located near $E_{\rm F}$ for $E=E_0$.

    Although the present theoretical model has three magnetic MLs, the
  increase in Fe layers induces the in-plane component of the magnetic
  anisotropy due to the magnetic dipole–dipole interaction in the
  two-dimensional alignment of the magnetic moments\cite{Szunyogh1995}. Taking into
  account such magnetic anisotropy, which is insensitive to the EF, the
  total MAE can be reduced to a small positive energy ($\sim0.6$ mJ/m$^{2}$,
  assuming an Fe layer 1.5 nm thick)\cite{Szunyogh1995},) such as the observed MAE
  ($0.31$ mJ/m$^{2}$)\cite{Nozaki2013}.

  In our model system, the negative slope of the MAE variation
  at a finite negative EF is understood to represent electron leakage
  from the interface to the MgO(R) layer, particularly to the surface
  adjacent to the vacuum layer. In the real system\cite{Nozaki2013}, the existence of
  charging spots (places where electrons are trapped) in the MgO(R)
  layer may be needed to explain the observed negative bias
  voltage. There could be an impurity site or a defect site in such
  systems. A probable site of origin is the B or O element. The
  existence of charging spots is consistent with the fact that the
  variation in the NOE is reduced on the interface at EFs below $E_{\rm c}$. The
  number of negative charges at the charging spots increases as the
  external EF decreases, reducing the effective EF imposed on the
  Fe/MgO(R) interface. In addition, another
  candidate for charging spots
  is an IRS at another MgO/Fe interface. This IRS should be assumed not
  to contribute to electron conduction along the perpendicular
  direction. In the previous experiment\cite{Nozaki2013}, there is an Fe-alloy/MgO/Fe
  junction, in which the IRS at MgO/Fe\cite{Butler2001} may act as a charging spot
  when negative voltages are applied, supposing that the IRS forms a set
  of localized states (nonconducting states).

  As shown in Fig. 2, the EF at which the inverse variation occurs is shifted from zero EF in
  the theoretical approach, in contrast to the experiment. This feature
  may be attributed to the difference between the model and real
  systems, in particular the thickness of the magnetic layer sandwiched
  by MgO and the magnetic material itself.  This is inferred from the
  fact that the reverse point of $E_{\rm c}$ should depend on the details of the
  electronic states around $E_{\rm F}$. If the magnetic layer becomes thicker,
  that is, the number of electrons increases (as for $E < 0$), the
  substantial $E_{\rm F}$ for the entire system may shift from $E_0$ toward $E_{\rm c}$. The
  lack of theoretical data for $E < E_{1}$ is related to an inaccurate energy
  position of the O $2p$ level (less binding) at the interfaces, which is
  improved by an advanced electronic structure calculation approach,
  such as the quasiparticle selfconsistent GW approximation\cite{Kotani2007}. These
  problems might be solved in future studies for the development of
  EF-driven magnetic devices.

    In summary, we studied the effect of an
  EF on the interface magnetic anisotropy in the double interface system
  MgO(L)/ Fe/MgO(R). A first-principles electronic structure calculation
  indicated that nonlinear behavior of the MAE was the intrinsic feature
  of such a magnetic layer with an interface.  The variations of the MAE
  with the EF are in good agreement with the experimental values both
  for $E < E_{\rm c}$ and $E > E_{\rm c}$. Our theoretical approach revealed that, under
  electron depletion, the decrease in the NOE at the Fe in Fe/ MgO(R)
  was enhanced by the existence of the MgO(L)/Fe interface, whereas
  under electron accumulation, an electron leak occurred on the Fe in
  Fe/MgO(R), leading to an increase in the MAE. The electronic
  structures indicate that such a leak is also intrinsic in the magnetic
  layer with the interface.  The theoretical base obtained here greatly
  encourages experimental research on such nonlinear behavior to develop
  new EF-driven devices that use the MAE.
  
Acknowledgements \\
The authors thank T. Nozaki and M. Otani for stimulating discussions. The
  computation in this work was performed using the facilities of the
  Supercomputer Center, Institute for Solid State Physics, University of
  Tokyo. This work was supported in part by Grants-in-Aid for Scientific
  Research from JSPS/MEXT (Grant Nos. 22104012 and 23510120) and the
  Computational Materials Science Initiative (CMSI), Japan.


\begin{thebibliography}{99}
\bibitem{Ohno2010}
H. Ohno: Nat. Mater. {\bf 9} (2010) 952.

\bibitem{Ohno2000}
H. Ohno, D. Chiba, F. Matsukura, T. Omiya, E. Abe, T. Dietl, Y. Ohno, and K. Ohtani: Nature {\bf 408} (2000) 944.

\bibitem{Weisheit2007}
M. Weisheit, S. F${\rm \ddot{a}}$hler, A. Marty, Y. Souche, C. Poinsignon, and D. Givord: Science {\bf 315} (2007) 349.

\bibitem{Maruyama2009}
T. Maruyama, Y. Shiota, T. Nozaki, K. Ohta, N. Toda, M. Mizuguchi, A. A. Tulapurkar, T. Shinjo, M. Shiraishi, S. Mizukami, Y. Ando, and Y. Suzuki: Nat. Nanotechnol. {\bf 4} (2009) 158.

\bibitem{Shiota2012}
Y. Shiota, T. Nozaki, F. Bonell, S. Murakami, T. Shinjo, and Y. Suzuki: Nat. Mater. {\bf 11} (2012) 39.

\bibitem{Naganuma2011}
H. Naganuma, L. Jiang, M. Oogane, and Y. Ando: Appl. Phys. Express {\bf 4} (2011) 019201.

\bibitem{Nozaki2013}
T. Nozaki, K. Yakushiji, S. Tamaru, M. Sekine, R. Matsumoto, M. Konoto, H. Kubota, A. Fukushima, and S. Yuasa: Appl. Phys. Express {\bf 6} (2013) 073005.

\bibitem{Suzuki2011}
Y. Suzuki, H. Kubota, A. Tulapurkar, and T. Nozaki: Phil. Trans. R. Soc. A, {\bf 369} (2011) 3658.

\bibitem{Niranjan2010}
M. K. Niranjan, C.-G. Duan, S. S. Jaswal, and E. Y. Tsymbal: Appl. Phys. Lett. {\bf 96} (2010) 222504.

\bibitem{Nakamura2010}
K. Nakamura, T. Akiyama, T. Ito, M. Weinert, and A. J. Freeman: Phys. Rev. B {\bf 81} (2010) 220409 (R).

\bibitem{Hotta2013}
K. Hotta, K. Nakamura, T. Akiyama, and T. Ito: J. Korean Phys. Soc. {\bf 63} (2013) 762.

\bibitem{Miyajima2009}
T. Miyajima, T. Ibusuki, S. Umehara, M. Sato, S. Eguchi, M. Tsukada, and Y. Kataoka: Appl. Phys. Lett. {\bf 94} (2009) 122501.

\bibitem{Oda2005}
T. Oda and A. Hosokawa: Phys. Rev. B {\bf 72} (2005) 224428.

\bibitem{Perdew1992}
J. P. Perdew, J. A. Chevary, S. H. Vosko, K. A. Jackson, M. R. Pederson, D. J. Singh, and C. Fiolhais: Phys. Rev. B {\bf 46} (1992) 6671.

\bibitem{Otani2006}
M. Otani and O. Sugino: Phys. Rev. B {\bf 73} (2006) 115407.

\bibitem{Tsujikawa2009}
M. Tsujikawa and T. Oda: Phys. Rev. Lett. {\bf 102} (2009) 247203.

\bibitem{Tsujikawa2011}
M. Tsujikawa, S. Haraguchi, T. Oda, Y. Miura, and M. Shirai: J. Appl. Phys. {\bf 109} (2011) 07C107.
	
\bibitem{Tusche2006}
\textcolor{black}{C. Tusche, H. L. Meyerheim, N, Jedrecy, G. Renaud, and J. Kirschner: Phys. Rev. B {\bf 74} (2006) 195422.}

\bibitem{Shimabukuro2010}
R. Shimabukuro, K. Nakamura, T. Akiyama, T. Ito: Physica E {\bf 42} (2010) 1014.

\bibitem{Ikeda2010}
S. Ikeda, K. Miura, H. Yamamoto, K. Mizunuma, H. D. Gan, M. Endo, S. Kanai, J. Hayakawa, F. Matsukura, and H. Ohno: Nat. Mate. {\bf 9} (2010) 721.

\bibitem{Nozaki2010}
T. Nozaki, Y. Shiota, T. Shinjo, and Y. Suzuki: Appl. Phys. Lett. {\bf 96} (2010) 022506.

\bibitem{Endo2010}
M. Endo, S. Kanai, S. Ikeda, F. Matsukura, and H. Ohno: Appl. Phys. Lett. {\bf 96} (2010) 212503.

\bibitem{Tsujikawa2012}
\textcolor{black}{M. Tsujikawa, S. Haraguchi, and T. Oda: J. Appl. Phys.  {\bf 111} (2012) 083910.}

\bibitem{Nakamura2009}
K. Nakamura, R. Shimabukuro, Yuji Fujiwara, and T. Ito: Phys. Rev. Lett. {\bf 102} (2009) 187201.

\bibitem{Bruno1989}
P. Bruno: Phys. Rev. B {\bf 39} (1989) 865.

\bibitem{Zermatten2008}
P.-J. Zermatten, G. Gaudin, G. Maris, M. Miron, A. Schuhl, C. Tiusan, F. Greullet, and M. Hehn: Phys. Rev. B {\bf 78} (2008) 033301.

\bibitem{Szunyogh1995}
L. Szunyogh, B. \'Ujfalussy, and P. Weinberger: Phys. Rev. B {\bf 51} (1995) 9552. 

\bibitem{Butler2001}
W. H. Butler, X.-G. Zhang, and T. C. Schulthess, andJ. M. MacLaren: Phys. Rev. B {\bf 63} (2001) 054416.

\bibitem{Kotani2007}
T. Kotani, M. van Schilfgaarde, and S. V. Faleev: Phys. Rev. B {\bf 76} (2007) 165106.
\end{thebibliography}
\end{document}